\documentclass{article}
\usepackage[dvips]{graphicx}
\graphicspath{{images/}}
\usepackage{epsfig}
\usepackage{longtable}
\begin{document}

\author{             V.Blinov and V.L.Golo\\
             Department of mechanics and mathematics, \\
             the Lomonosov Moscow State University, Moscow, Russia
}

\title{Hyper-sound as a means for generating inter-strand defects in a duplex of the DNA}

\date{August 7, 2009}

\maketitle

\begin{abstract}
    The formation of  bubble defects of the double stranded DNA is
    treated according to the Lifshits theory of disordered chains.  
    A molecule of the DNA is modelled on a harmonic lattice with 
    nearest neighbour interaction, elastic constants being randomly distributed.
    The helicoidal symmetry is accommodated through a chiral field at sites of the
    lattice.   The number of sites varies from $100$ to $300$,   corresponding to  DNA segmnets  
    of persistence length. We find  
    the spectra of elastic eigen-modes that  mimic inter-strand excitations of the duplex. 
    The  frequency distribution shows peaks and valleys 
    at the high-frequency end of the spectra,  in accord with the general theory.
    External excitations may lead to a parametric resonance 
    that can generate localized modes of the lattice.
    In real life pumping hyper-sound may generate a  resonance similar to that 
    studied in this paper, and    thus result in excitation of inter-strand modes
    and possible formation of bubbles, in the duplex of the DNA.   
\end{abstract}

\section{ Introduction}

There has been  remarkable progress in  the
studies~\cite{Bonnet1}--- \cite{Leroy} of defects of the DNA due to
breaking hydrogen bonds and formation of local
denaturation zones, or bubbles. It is generally accepted that the  phenomenae are important for
understanding basic facts of molecular biology, i.e. denaturation,
replication, transcription. A number of recent papers have been
dedicated to working out a theoretical description of the
phenomenon, see paper~\cite{hanke} and references therein.
Generally, the dynamics of bubble formation has been considered
either within the stochastic, or
nonlinear mechanics, see
\cite{hanke}. In papers  \cite{Hennig_1},\cite{Hennig_2}, \cite{Wattis}  the authors
suggest to take into account  irregular structure of the DNA.
In fact, the conformation of the DNA as regards positions of
base-pairs and values of elastic constants describing its
elastic properties, suffers serious deviations from the picture of
ideal double-helix~\cite{Hassan}. The problem is to what extent the
irregularities of duplex may be involved in the formation of
inter-strand defects of the DNA.

\begin{figure}[h!]
\centering
        {
            \includegraphics[height=6cm, keepaspectratio]{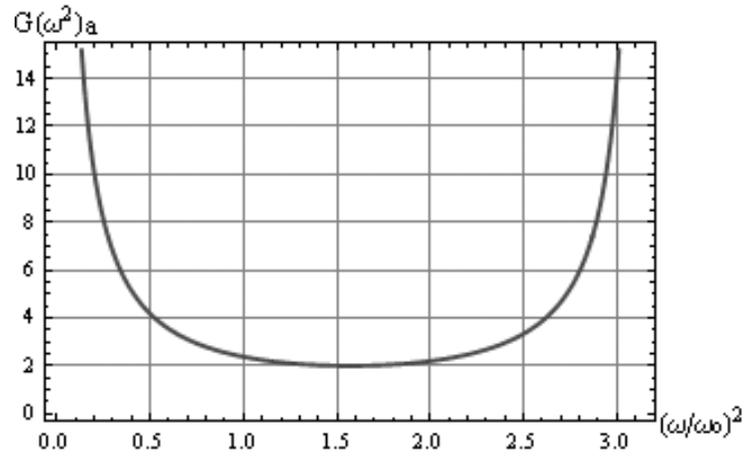}
            \caption{ Squared frequency spectra for regular one dimensional  lattice .}
            \label{fig:g-omega2}
        }
\end{figure}

\begin{figure}[h!]
\centering
        {
            \includegraphics[height=6cm, keepaspectratio]{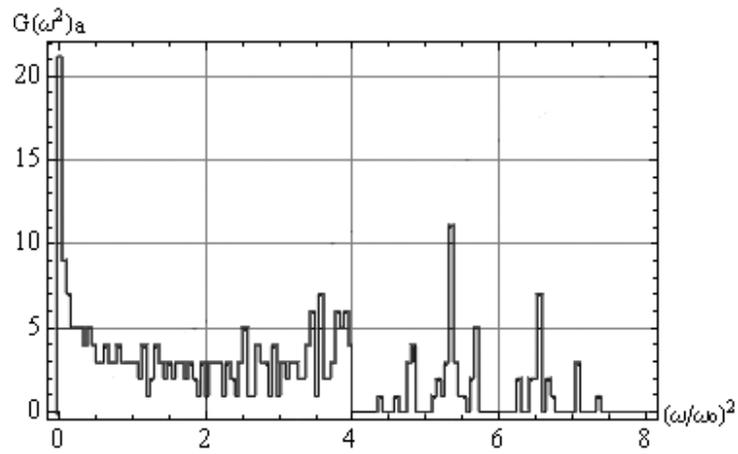}
            \caption{ Squared frequency spectra for disordered lattice.
                      Elastic constants selected randomly from 1 and 3.}
            \label{fig:g-omega2dist}
        }
\end{figure}

The answer may be provided by  the theory of dynamic
excitations in random media worked out, years ago , by I.M.Lifshits
and his colleagues,~\cite{Lifschitz1} --- \cite{Lifschitz5} (see
also  \cite{Montroll}), \cite{Ziman}). According to Lifshits' theory
defects of media serve centers for the localization of elastic excitations, see
\cite{Lifschitz1} --- \cite{Lifschitz5}. It should be noted that the introduction of disorder 
results in drastic changes in the shape of spectra, 
as is illustrated in Figs.\ref{fig:g-omega2}~---\ref{fig:g-omega2dist}.

One can synthesise molecules of the DNA with a prescribed sequence of base-pairs,
disordered or otherwise, and thus obtain a  variety of  elastic systems having elastic properties
required  for generating various frequency spectra. Thus, one may expect that the DNA may serve a means
for experimental study of chaotic elastic dynamics.

 In this respect one may suggest that
 the formation of bubbles in a molecule of the DNA
is due to inter-strand motions  acquiring an amplitude large enough 
for the  hydrogen bonds of a molecule to be broken. Of course, the actual process demands 
taking into account non-linear efects. But the later problem is hard to be solved, presently.  
One may only suggest that generating inter-strand modes of large amplitude, for example in a resonance regime,
is a sign that the bubbles run a chance to appear.  Therefore, it is worthwhile 
to look for localzied inter-strand modes in the dynamics of disordered chains corresponding to a molecule of the DNA.

\begin{figure}[h!]
\centering
        {
            \includegraphics[height=6cm, keepaspectratio]{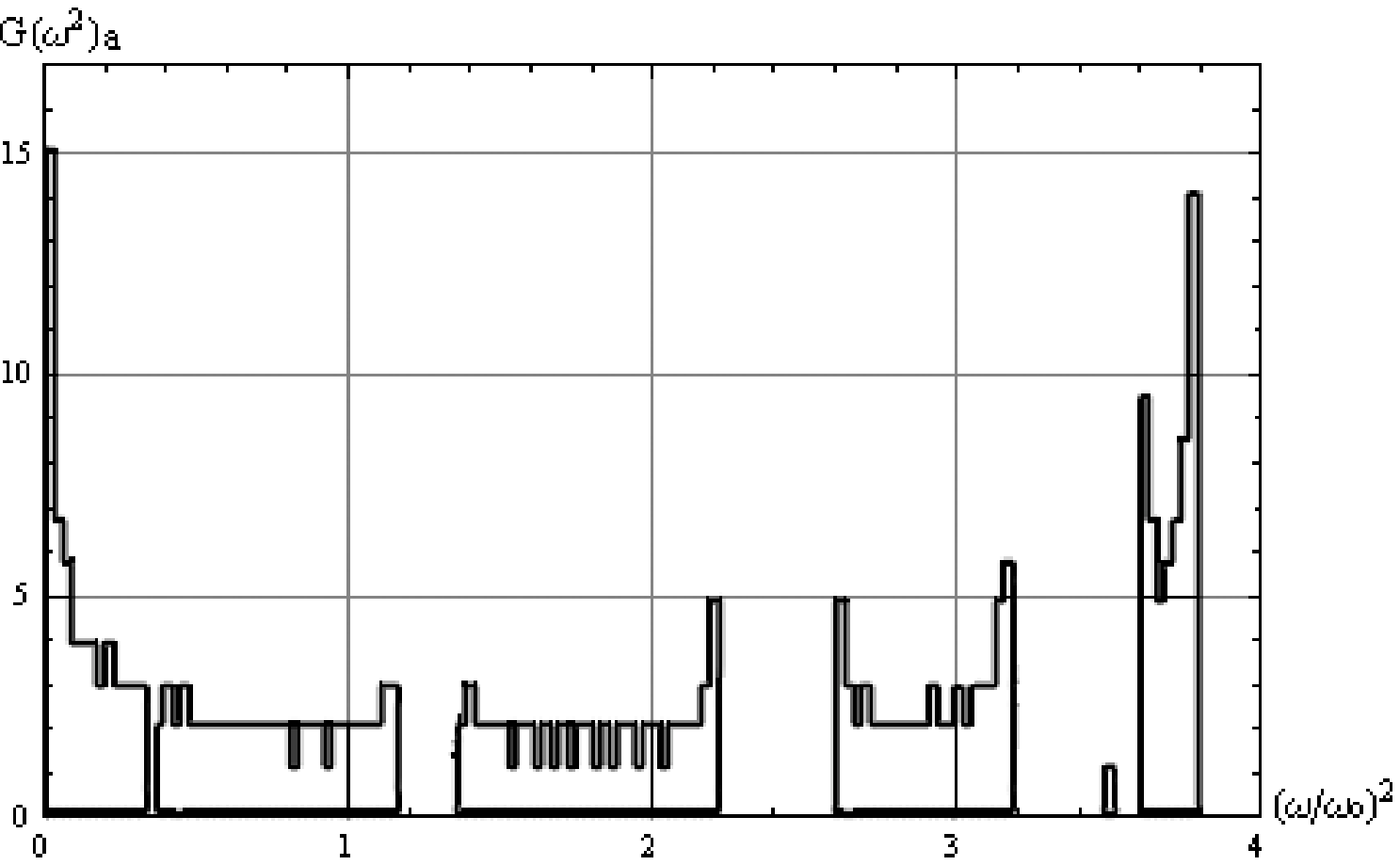}
            \caption{ Squared frequency spectra for periodic lattice.
                      Elastic constants are equal, masses form
                      sequence  ...-1-1,5-1-1-1-1-1,5-1-1-1...}
            \label{fig:g-omega2per}
        }
\end{figure}

The study of spectra of disordered chains, or lattices, has
followed mainly two paths.  A number of papers employed analytical
methods, which generally could provide only qualitative description
of the spectra, see \cite{Montroll}.  It was P.Dean, \cite{Dean},
who developed a powerful computer technic for analysis situs of  the
frequency distribution of disordered lattices. 
In this paper we use  the algorithms worked out by Dean for the
analysis  of inter-strand modes of the DNA  of a simple 
qualitative model which mimicks the most essential conformational properties of the DNA.

\section{Resonance in random helicoidal lattice.}

\begin{figure}[h!]
\centering
        {
            \includegraphics[height=5cm, width=7cm]{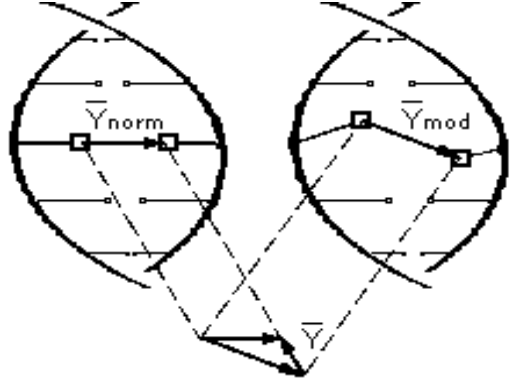}
            \caption{ Site $n$ corresponding to a base-pair.
                      Relative position of the bases  denoted by
                      vector $\vec{Y}_{norm}$. Vector
                      $\vec{Y}_{mod}$  indicates relative positions
                      of the base-pairs in the deformed molecule.
                      Vector $\vec{Y} = \vec{Y}_{mod} -
                      \vec{Y}_{norm}$ describes the dynamics of the
                      base-pair within the framework of the model.
        }
            \label{fig:dnaYIntro}
        }
\end{figure}

\begin{figure}[h!]
\centering
        {
            \includegraphics[height=6cm, width=9cm]{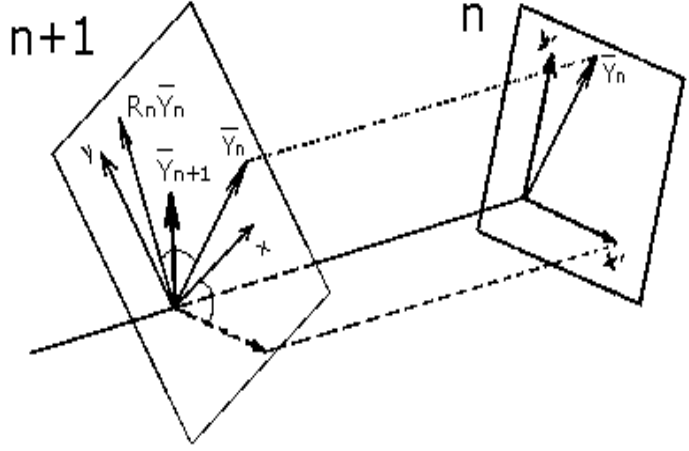}
            \caption{ Local frames due to the helical symmetry.
            Coordinate systems $x-y$ and $x^{\prime}-y^{\prime}$, corresponding to sites $n$ and
            $n+1$, transformed into each other by rotation $R_n$.
            The change in values of vectors $\vec Y_n$ and $\vec Y_{n+1}$
            allows for  rotation transformation $R_n$, according to
            expression $\vec R_{n+1} - R_n  \cdot  \vec Y_n$.
            Deviations from the equilibrium position coincide with those
            corresponding to a helicoidal lattice formed by vectors $\vec Y_n$.
            In what follows $R_n = R$.}
            \label{fig:planes}
        }
\end{figure}

It is extremely difficult, if at all possible, directly to employ the theory of disordered media
indicated above for the needs of the DNA.  Therefore, one has to
resign to  realistic simplifications of the latter.
In building appropriate models we are to take into account the
following basic features:
\begin{enumerate}
    \item  a molecule of the DNA consists of two strands;
    \item  it has helicoidal symmetry;
    \item  its base-pairs  generally form a random sequence.
\end{enumerate}
The double-stranded structure of the DNA provides an opportunity for
its base-pairs' changing  relative positions in space, see
Fig.\ref{fig:dnaYIntro}. This fact is of primary importance for
describing the internal, or inter-strand, dynamics of the DNA, in which strands of 
a molecule move from their equilibrium positions. If the amplitudes of the motion are small
enough the duplex structure of the molecule is preserved, even though deformed. 
The  two strands are still  joined by hydrogen bonds, 
their relative positions suffering only small changes.
But there is a need for a more quantitative description of this phenomenon.
Aiming at a simple picture of a molecule of the DNA, we choose the
{\em one dimensional interpretation}  as a 1D lattice whose sites $n
= 1, 2, \ldots , N$ denote base-pairs of the molecule, and describe
the change in relative positions of base-pairs with vectors $\vec
Y_n$. Thus, we obtain a very crude picture of a molecule of the DNA,
which still preserves the essence of two strands and brings forward
the helicoidal dynamics of the system~\footnote{One may recall the
familiar story "Dancing Men" by Arthur Conan Doyle, in which crude
pictures of human beings served a code for the plot. }. It is worth
noting that $1D$ structure of the  model is in agreement with the
conformation of a molecule of the DNA of length less than the
persistent one, that is $300 - 1000 \AA$, depending on ion strength
of an ambient solvent. We note that a  model of the DNA that
corresponds to a discreet form of non-linear Klein-Gordon equation and 
take into account irregularities of the molecule,
is put forward in paper \cite{Wattis}, but it does not take into account 
the helicoidal structure of the DNA.
In this paper, following \cite{gky}, 
we accommodate the helicoidal structure of the
duplex with the help of a gauge field
describing the helicoidal, or helix,  structure~\footnote{The concept is quite
familiar in gauge field theories, see H.Kleinert, Gauge Fields in
Condensed Matter, World Scientific, Singapore (1989), and is
essentially the same as the angular velocity  of a top. It dates
back to Poisson's time and the theory of motion of rigid bodies, see
E.J.Routh, Dynamics of a System of Rigid Bodies, Ch. V, Macmillan,
London (1892). Similar considerations are employed in deriving the Kirchhoff
equations for the motion of a rigid body in ideal fluid, see H.Lamb,
Hydrodynamics, Cambridge University Press, London(1975).} . To that end
we consider a local frame at each site of the lattice, which provides a reference for local  strain 
caused by displacements of {\em adjacent base-pairs}. The
frames are rotated by matrix R due to the transition from  cite $n$
to site $n+1$, and by the inverse matrix $R^{-1}$ for the inverse
transition from $n+1$ to $n$ as is illustrated in
Fig.\ref{fig:planes}. A deviation from the rule results in straining
the system and increasing the elastic energy.
To provide an analytical framework for the above  picture
we assume that the kinetic energy  is related to the dynamics of the
vectors $\vec Y_n$ and can be cast in the form
$$
    T = \sum\limits_{n=1}^N
            \left[ \frac{m_n}{2} ( \stackrel{.}{\vec Y_n })^2
            \right ]
$$
in which masses $m_n$ describes inertial effects accompanying 
the relative motion of bases inside the pairs.
An expression for the potential energy should describe both the
energy related to mutual displacement of the strands, which we cast
in the form
$$
    U_{dev} = \sum\limits_{n=1}^N
            \frac{\epsilon_n}{2} ( \vec Y_n )^2
$$
and the energy caused by the rotation of adjacent base-pairs, which
we may write down as follows
$$
    U_{rot} = \sum\limits_{n=1}^N \frac{k_n}{2}
            \left( \vec R_{n+1} - R \cdot \vec Y_n \right )^2
$$
The total potential energy $U$ is equal to
$$
    U = U_{rot} + U_{dev}
$$
It is worth noting that $U_{dev}$ equals to zero if $\vec Y_n = 0,
\quad n = 1,2, \ldots, N$, as well as  $U_{rot} = 0$, when vectors
$\vec Y_{n+1}$ and $\vec Y_n$ are transformed into each other by
rotation $R$.
\begin{figure}[h!]
\centering
        {
            \includegraphics[width=15cm, keepaspectratio]{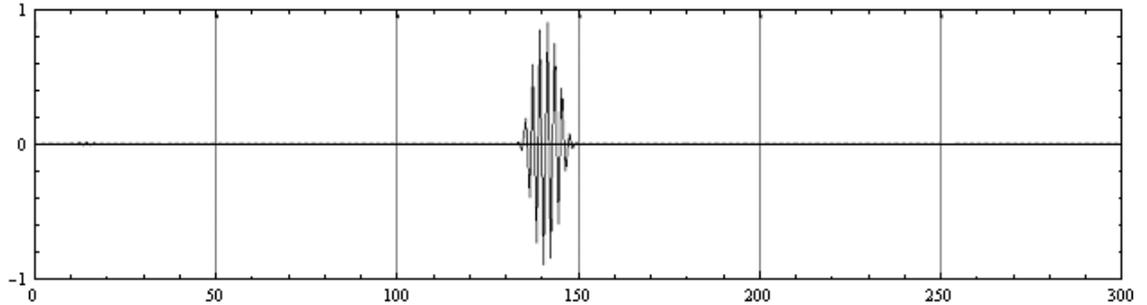}
            \caption{ Localized excitation due to a parametric
            resonance. All the masses equal to 1. Elastic constants
            are 1 and 1.5 distributed with equal probabilities.
            Localization sites 139 - 148, the excitation's frequency
            5.63.
        }
        \label{fig:param_loc}
        }
\end{figure}

\begin{figure}[h!]
\centering
        {
            \includegraphics[width=15cm, keepaspectratio]{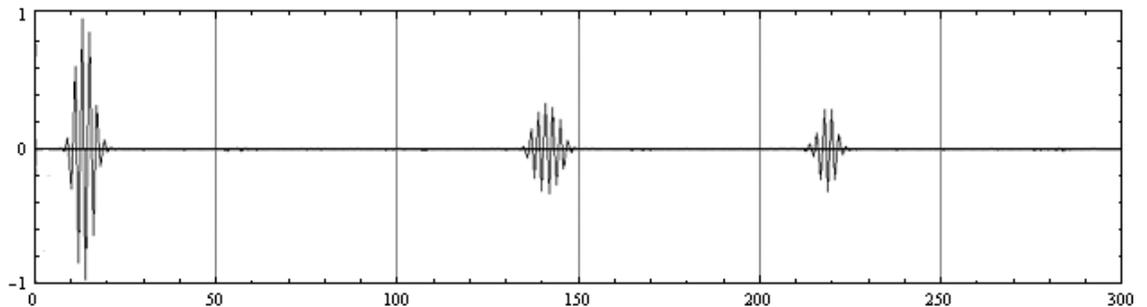}
            \caption{ Localized excitation with frequency 5.40.
            Parameters of the lattice are the same as in
            Fig.\ref{fig:param_loc}.
        }
        \label{fig:param_mul}
        }
\end{figure}

Thus, we obtain the Lagrange function of our system
$$
    L = T - U
$$
where the potential energy $U$ is given by the expressions indicated
above. The equations of motion read as usual, and, to avoid
unnecessary formulae, we shall not write  them down  explicitly.
It is important that the equations of motion can be cast in a more
tractable form. To that end we resort to 
a simple arithemetic device by using complex notations. The point is that 
vectors $\vec Y_n$ being perpendicular to the axis of the molecule
we may consider them two-dimensional, $ \vec Y_n = (Y_1, Y_2)$ and
assign  complex number $z_n = Y_n^1 + i Y_n^2$ to vector $\vec Y_n$.
Then the rotation given by matrix $R$ introduced above corresponds
to the multiplication by complex number $    e^{i \theta}, \theta =
\pi / 5   $. Let us introduce new variables $y_n$ defined  by the
equation
\begin{equation}
    y_n = e^{- i \Theta} \, z_n, \quad \Theta = (n - 1) \, \theta, \quad
    n = 1,2, \ldots , N -1
    \label{complex}
\end{equation}

\begin{figure}[h]
    \includegraphics*[width=14cm, keepaspectratio]{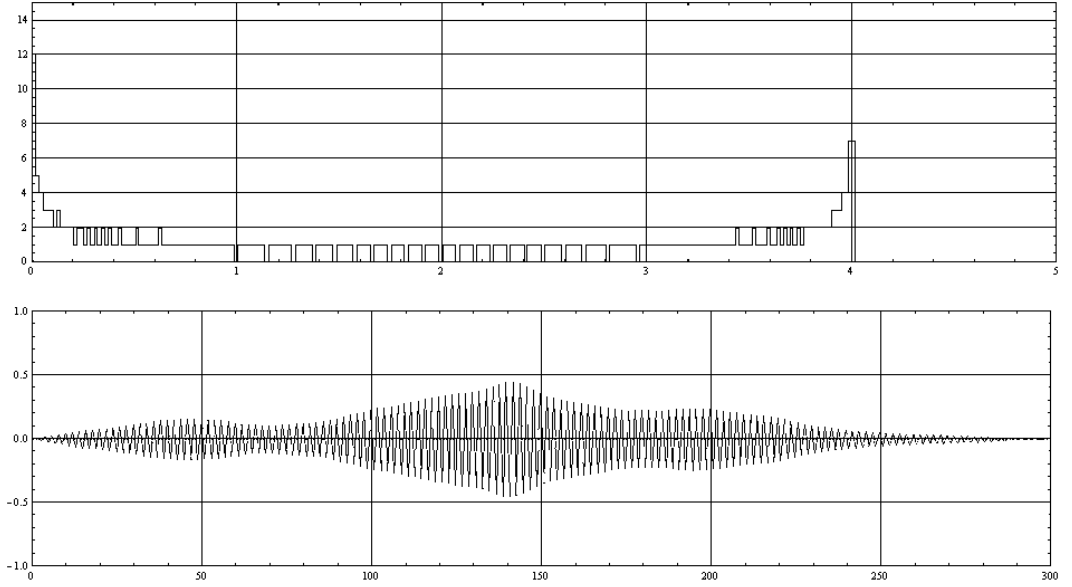}
    \caption{\small{Squared eigenvalues density $G(\omega^2)$ histogram for a chain with $\delta=0.5$,
    $\epsilon=0.003$ below normal mode corresponded to eigenvalue 4.006082.}}
\end{figure}
\begin{figure}[h]
    \includegraphics*[width=14cm, keepaspectratio]{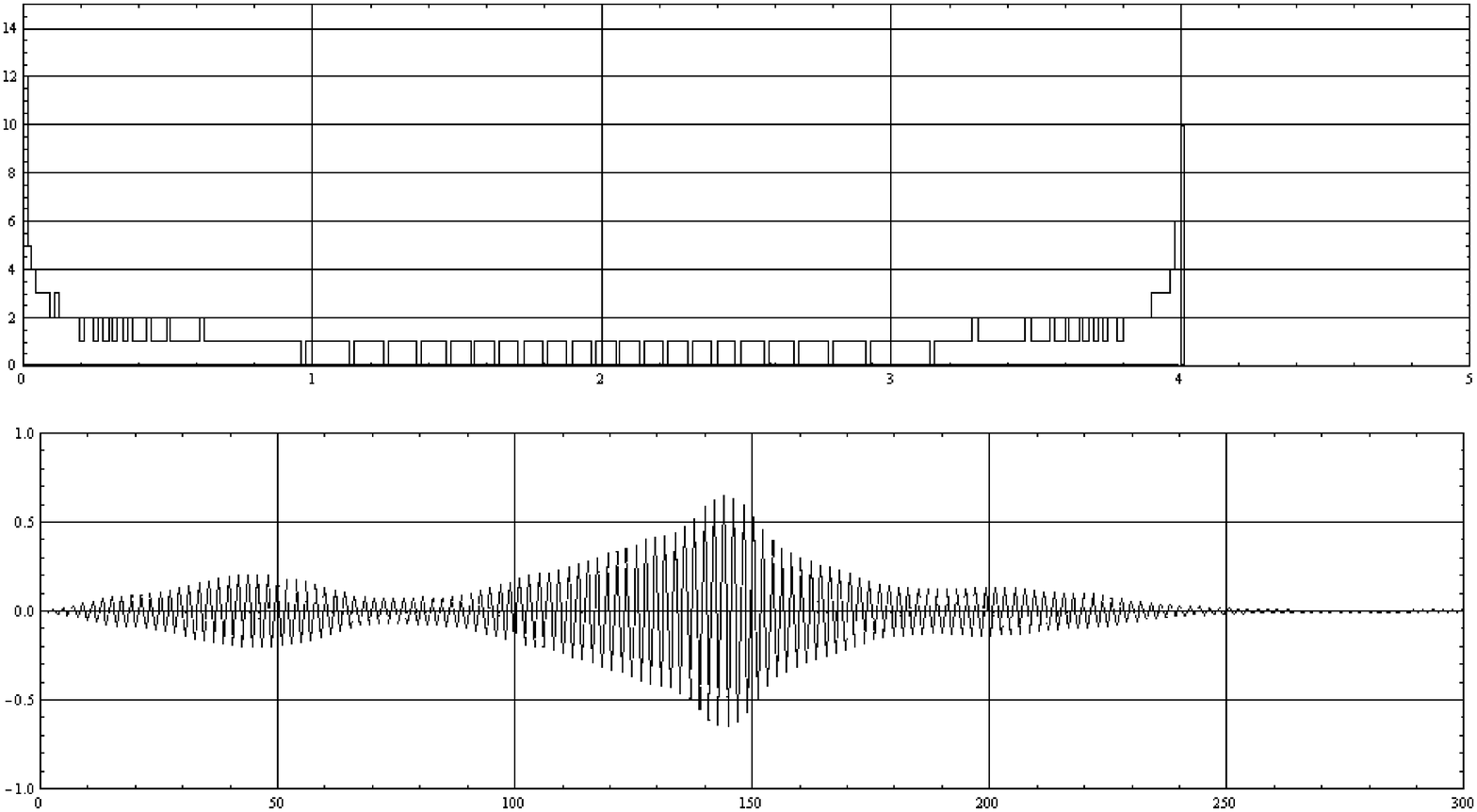}
    \caption{\small{Squared eigenvalues density $G(\omega^2)$ histogram for a chain with $\delta=0.5$,
    $\epsilon=0.06$ below normal mode corresponded to eigenvalue 4.012972.}}
\end{figure}
\begin{figure}[h]
    \includegraphics*[width=14cm, keepaspectratio]{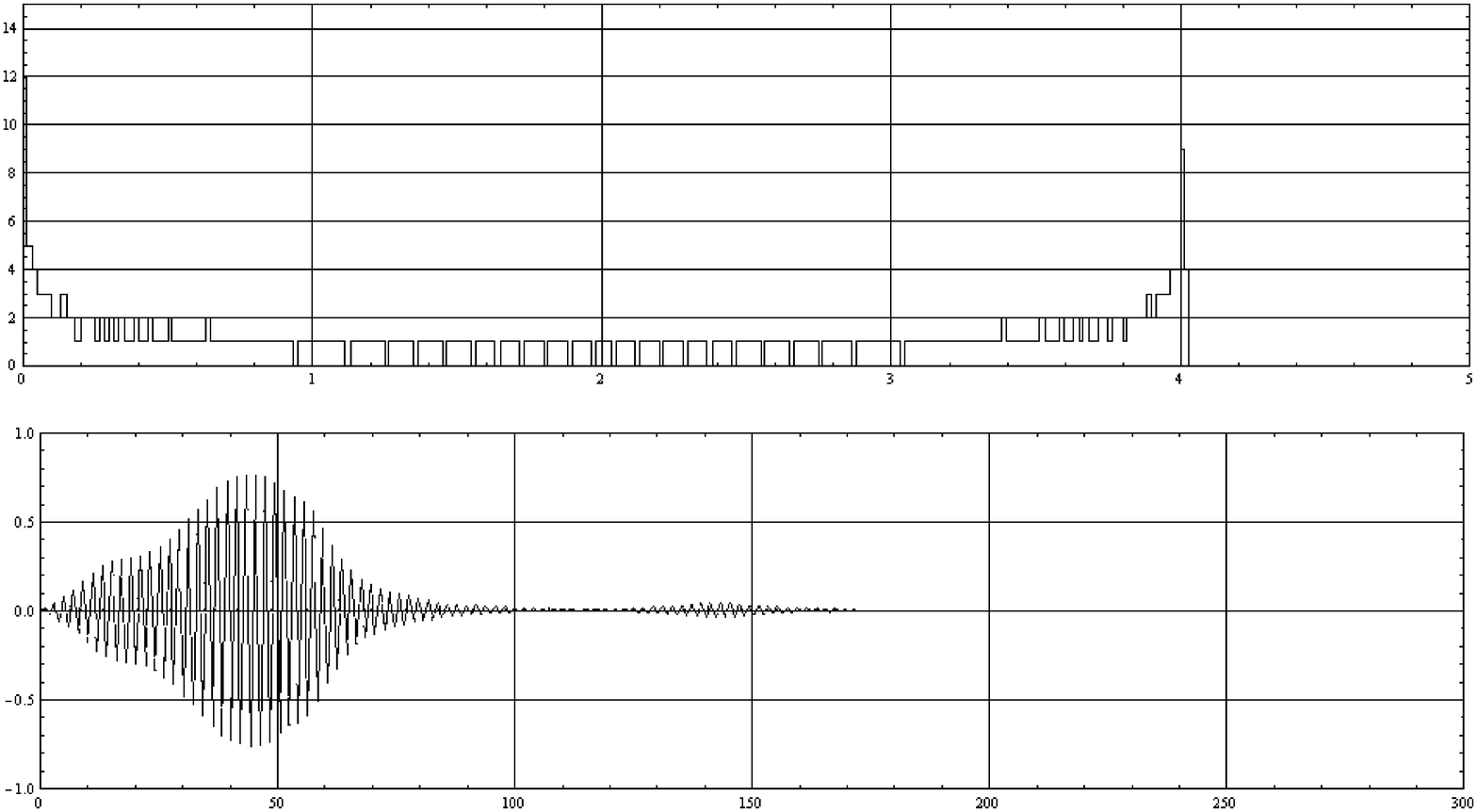}
    \caption{\small{Squared eigenvalues density $G(\omega^2)$ histogram for a chain with $\delta=0.5$,
    $\epsilon=0.009$ below normal mode corresponded to eigenvalue 4.020306.}}
\end{figure}
\begin{figure}[h]
    \includegraphics*[width=14cm, keepaspectratio]{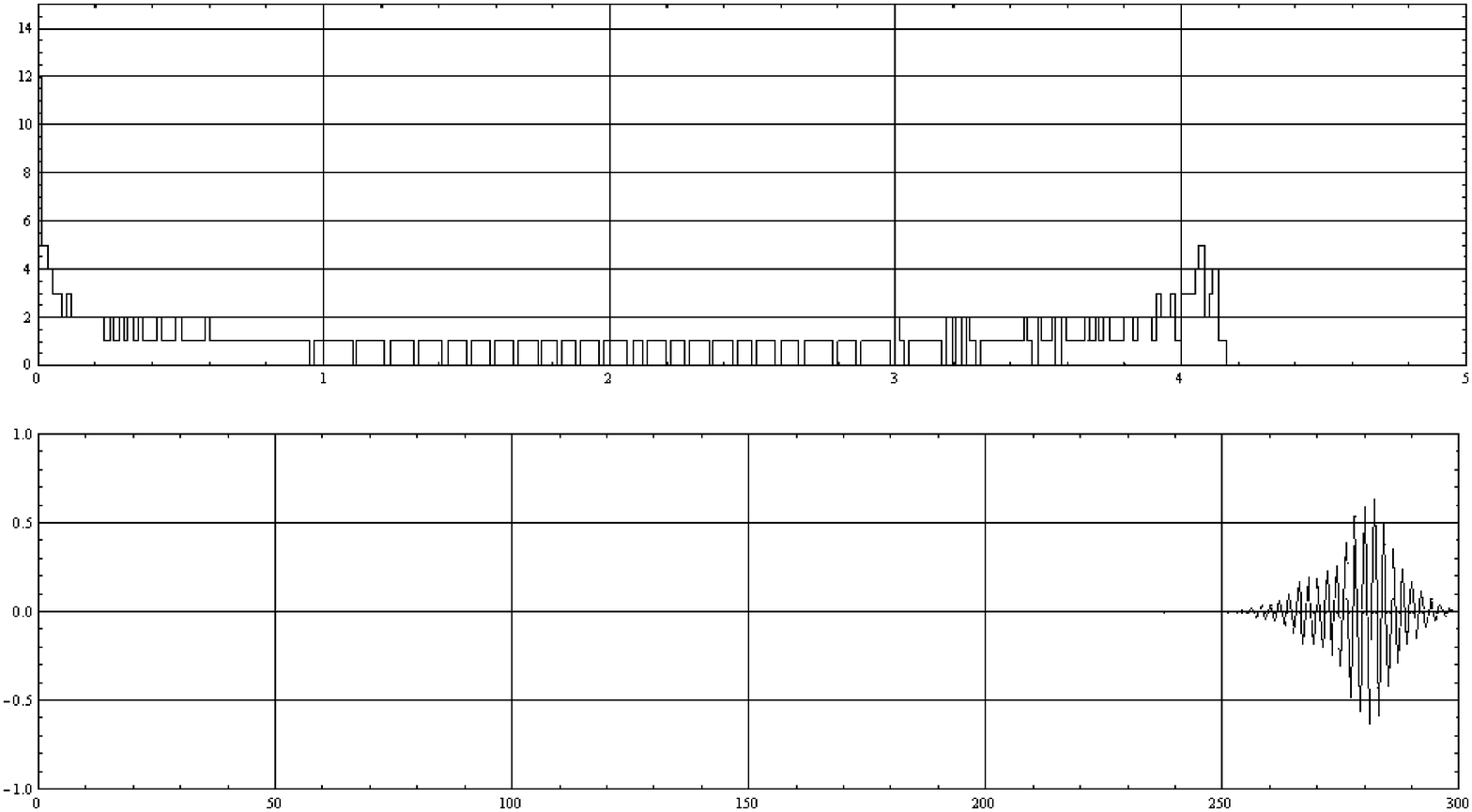}
    \caption{\small{Squared eigenvalues density $G(\omega^2)$ histogram for a chain with $\delta=0.5$,
    $\epsilon=0.05$ below normal mode corresponded to eigenvalue 4.124748.}}
\end{figure}
\begin{figure}[h]
    \includegraphics*[width=14cm, keepaspectratio]{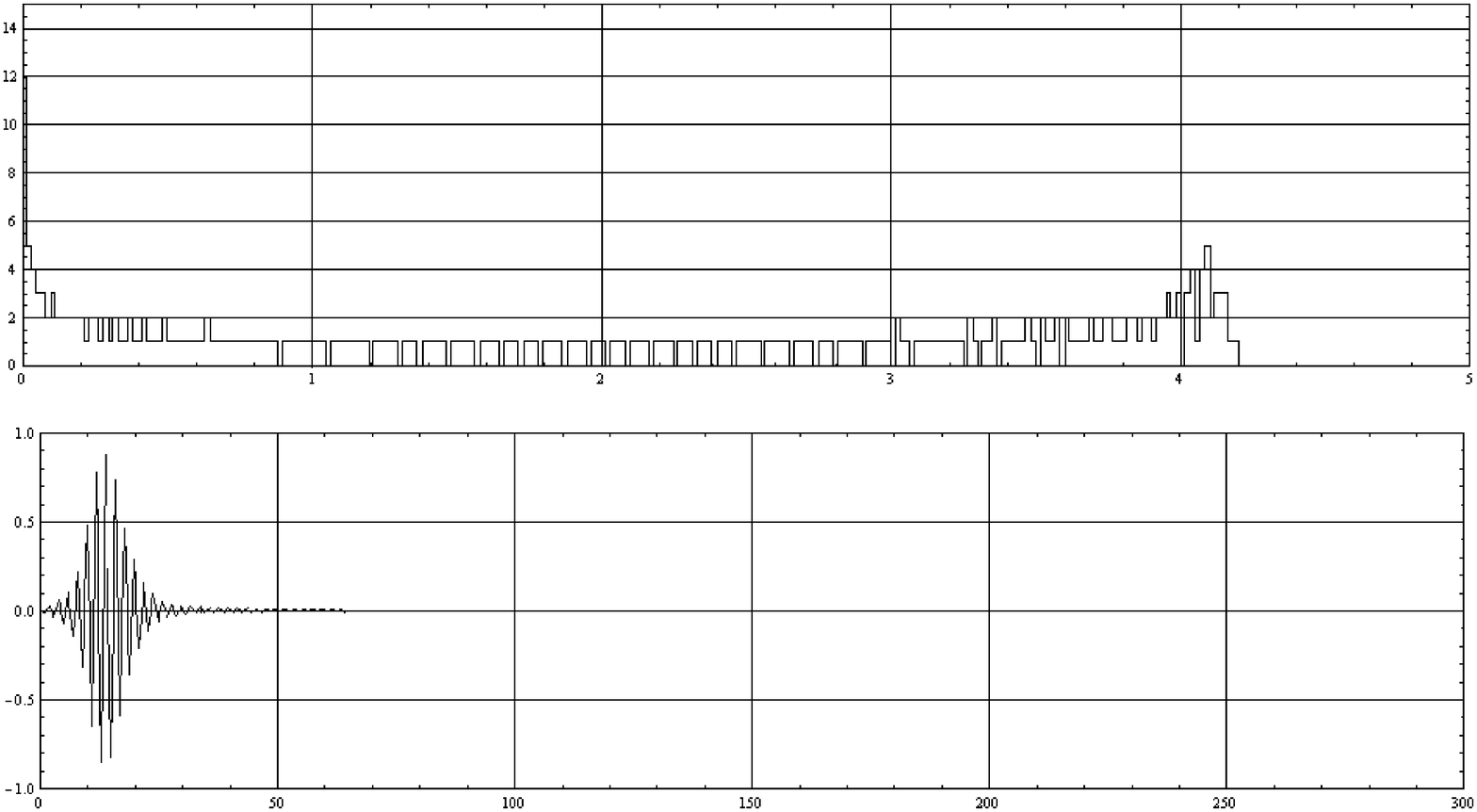}
    \caption{\small{Squared eigenvalues density $G(\omega^2)$ histogram for a chain with $\delta=0.5$,
    $\epsilon=0.06$ below normal mode corresponded to eigenvalue 4.168756.}}
\end{figure}

Using the $y_n$ we may cast the Lagrangian of our lattice in the
form
$$
L = \sum\limits_{n = 1}^N
        \left [
        \frac{m_n}{2} ( \dot{y}_n \cdot \dot{y}_n^{*}) -
        \frac{k_n}{2} \; \left | y_{n+1} - y_n \right |^2  -
        \frac{ \varepsilon_n}{2} \; | y_n |^2
        \right ]
$$
where $*$ signifies  complex conjugation, and 
the equations of motion for $y_n, \; \stackrel{*}{y}_n$ in the form
\begin{equation}
    m_i\ddot{y_i} =
    k_{i-1}y_{i-1}+k_iy_{i+1}-(k_i+k_{i-1})y_i-\varepsilon_iy_i
    \label{ModelBasic-1}
\end{equation}
and
\begin{equation}
    m_i\ddot{y_i}^* =
    k_{i-1}y_{i-1}^*+k_iy_{i+1}^*-(k_i+k_{i-1})y_i^*-\varepsilon_iy_i^*
    \label{ModelBasic-2}
\end{equation}

Thus we  split the equations of motion for
the helicoidal lattice into two complex equations
(\ref{ModelBasic-1}),(\ref{ModelBasic-2}) for a simple harmonic
lattice,  the chiral field given by matrix $R$ being accommodated by
substitution (\ref{complex}).  Now we may use the
method worked out by Dean, \cite{Dean}, for harmonic chains. The
fact that the chain variables, $y_n$ , are complex numbers is not
essential. Details of Dean's algorithm see in the
review article \cite{Dean}.
\begin{figure}[h!]
\centering
        {
            \includegraphics[width=10cm, keepaspectratio]{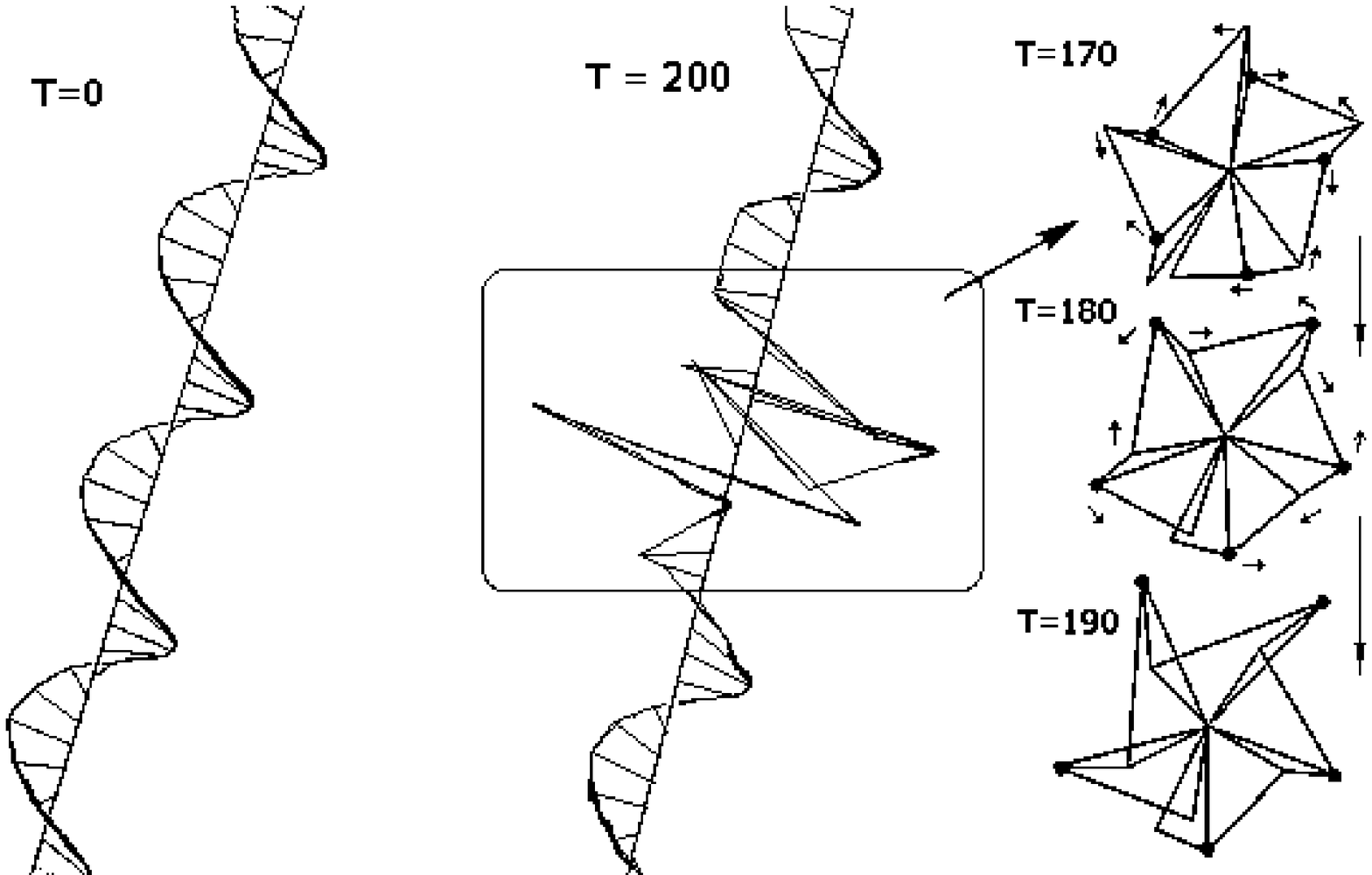}
            \caption{ 
             Picture on the left hand side illustrates the regular configuration.
             The central and the right ones show 
             localized excitations, frequency 5.63,
             localization at 139-148 base-pairs.}
            \label{fig:param_mul}
        }
\end{figure}

At this point it is necessary to indicate a feature important for the problem at hand.
It is generally accepted \cite{Ziman}, \cite{Dean}, \cite{Lifschitz1},   
that eigenmodes become localized
``even for a very small chaotic perturbation of the interaction potential'',\cite{Lifschitz1},p.39.  
The statement does not make any claims as to the size of localization, but our calculations indicate
that a magnitude of disorder given by values of elastic constants is essential.  We have made a series 
of calculations in which the elastic constants take on values $1$  or $1 - \epsilon$ with equal probability, and found that
depending on  $\delta$, the spectra of eigenmodes and the sizes of localization acquire the following shapes:
\begin{enumerate}
 \item $\delta = 0.001$ - distribution function does not change appreciably; no localization on the scale of
       the lattice;
 \item  $\delta = 0.006$ - distribution function does not change appreciably; no localization on the scale of
       the lattice;
 \item  $\delta = 0.009$ - distribution function looks the same but there is  localization on the scale of
       the lattice, see 
 \item  $\delta = 0.03$ - distribution deformed; localization  30 - 40 sites involved; 
 \item  $\delta = 0.05$ - distribution deformed; localization   30 - 40 sites involved; 
 \item  $\delta = 0.06$ - distribution deformed; localization within a small region.
\end{enumerate}
Therefore, we need to take into account values of random disturbance of elastic constants,  
if we are looking for localizations of a size less than that of the sample. In a situation pertinent to
a molecule of the DNA, it is reasonable to assume that elastic constants may have values of relative proportions 
1 and 1.5, in accord with the number of hydrogen bonds, 2 and 3, for the base-pairs A-T and G-C. It enables to obtain
fairly pronounced localizations confined to small regions of a lattice.  

The data and figures indicated above suggest that there is a large amount of
situations in which localized modes could be expected for duplexes of the DNA.
One may employ parametric excitations to generate them. 
Taking into account attenuating and damping effects we consider the dynamics described by the following equation
\begin{equation}
    m_i\ddot{y_i} =
    k_{i-1} y_{i-1}  +  k_iy_{i+1} - (k_i+k_{i-1}) y_i  
                     -  \varepsilon_i  y_iv  -   \gamma \dot{y}_i  +  F_i
    \label{Parametric}
\end{equation}
and choose external  excitation $F_i(t)$ given by
\begin{equation}
   \label{par_exc}
    F_i = 2 A \; y_i \; \sin{(2 \nu t)}
\end{equation}
It is important that there is a parametric resonance 
for frequency $\nu$ which is  the double of that of a lattice eigenmode.  The amplitude  $A$ of the
excitation is to be large enough to counterbalance dissipative effects due to the term $ \gamma \dot{y}_i$.
Since masses $m_i$ of the lattice sites, as well as dissipative and elastic constants, $\gamma_i$, and $k_i$,  
are randomly distributed.  It is difficult to write down an analogue of the Rayleigh 
condition, \cite{Rayleigh},  for the resonance
$$
	( \omega^2  -  \nu^2 )^2  =  A^2  -  \left( \frac{\gamma}{m}  \right)^2 \; \nu^2  
$$
If all the masses, as well the dissipative and elastic constants, are  equal, 
the above equation is valid for an eigenmode with frequency 
$\omega$.  At any rate, we may conclude that there is a threshold for 
amplitude $A$, below which one cannot obtain a resonance.

One may put forward the following arguments in favour of our approach.
The interactions within base-pairs being estimated, \cite{santa} , less $2 k_B T$,
thermal excitations of energy $k_B T$ are generally accepted as a cause of DNA breathing, i.e. opening and re-closing of hydrogen bonds,\cite{Bonnet1}, \cite{Bonnet2}. It is important that they admit an interpretation using wave theory.   L.I.Mandelstam, \cite{Mandelstam}, indicated that  waves of density introduced by  A.Einstein in the theory of  Brownian motion, \cite{Einstein}, are  elastic waves, so that one can consider density fluctuations as  a result of interference of the elastic waves, \cite{fabelinskii1}, \cite{fabelinskii2}, or hyper-sound.  In assessing the part  played by hyper-sound in the inter-strand dynamics it is necessary to compare characteristic sizes.  The diameter of a molecule of the DNA is approximately equal to $20 \AA$, the length of its segment comprising 10  base-pairs is $34 \AA$, whereas the wavelength of hyper-sound in the range of a few tens of $GHz$ is of order of several tens $\AA$, that is all characteristic sizes involved are of order $10^{-6} \div 10^{-7} \; cm$. 
The sizes are propitious for generating inter-strand motions.  Indeed, imagine a hyper-sound wave of wave-length approximately 
$40 \AA$, that is frequency $\propto 10^{12} \, Hz$.  The chances are that its crest may be located at one strand of the molecule 
whereas its bottom at the other. Therefore, there will emerge a torque and a strain, periodic in time, that pull apart the opposite strands.  If amplitudes of parts of the molecule are small enough it is only natural to suggest that the motion is to be described by terms linear in fields $\vec Y_n$.  Thus we arrive at the form (\ref{par_exc}) for the external excitation\footnote{Since we do not aim at studying the denaturation and other non-linear phenomena, we refrain from introducing  clever non-linear terms in the potential energy of a molecule of the DNA, and confine ourselves to harmonic approximation. }.

\section{CONCLUSIONS}

According to the arguments given above inter-strand defects, or bubbles, of a molecule of the DNA
could be due to an interplay between  random  sequences of base-pairs  and hyper-sound.  The latter is an agent that
generates density fluctuations, \cite{fabelinskii2} that are assumed to break hydrogen bonds.. Thus, we may suggest that 
pumping of hyper-sound  could make for  bubble defects, through  maintaining  inter-strand modes due to 
the parametric resonance effected by hyper-sound pumping. Whether this process really takes place depends on magnitude of dissipation
accompanying the inter-strand dynamics. Presently, it is often claimed that the attenuation is 
so strong that the inter-strand modes should be overdamped. 
The arguments are based on estimating relaxation  due to 
viscosity  due to surrounding liquid medium, \cite{Adair}. But in estimating the dynamics of 
liquid inside GHz-region one must take into account
that there the Navier-Stokes equations cease to be valid,\cite{fabelinskii2}. 
In this region one has to turn to the theory of GHz-hydrodynamics worked out by Mandelstam and 
Leontovic, \cite{fabelinskii1}, \cite{fabelinskii2}.  So far  there have been no estimates of 
the attenuation made within the framework of their theory. In this respect, it is worth noting that the Navier-Stokes theory
gives totally inadequate results for the treatment of the Mandelstam-Brilloiun light scattering due to hyper-sound.
In fact, there are different opinions as to the importance of the disspative effects:  
M.E.Davis and L.L. VanZandt, \cite{Zandt}, following  Maxwell's approach to hydrodynamics, came to the conclusion 
that the attenuation of inter-strand modes does not overdamp them;  
in paper \cite{ShihGeorghiou} the authors, on different grounds, come to the same conclusion.  Thus, the question 
of the attenuation of inter-strands is still open. Nonetheless, considerable efforts should be taken to circumvent 
undesirable effects of attenuation in experimental research. The important advaces in hyper-sound acoustics, in particular
constructing  transducers, \cite{Huynhn}, enables us to be reasonably optimistic as to the experimental feasibility
of hyper-sound investigation of the DNA. One may suggest that the attenuation of hyper-sound 
could be a means for detecting structural changes in the DNA.  It is worth noting that to that end, using orietationally organized structures of the DNA,
for example its liquid crystalline phases, \cite{Livolant}, would  be useful. Atomic force microscopy could be an important instrument as well for 
studying the formation of the bubbles, even more so as it permits to employ molecules of the DNA on solid substrate, and thus to
escape at least   difficulties due to the disspative  effects  as regards  molecules of the DNA in solution. For this argument we indebtted to K.V.Shaitan.

Finally, we would like to draw attention to the fact that a more refined but still practical model of the DNA,
for example that worked out in paper \cite{Kovaleva}, could make for better understanding chaotic dynamics of inter-strand modes, 
their bearing on physics of the  DNA, and possible biological implications.

We are thankful to K.V.Shaitan for the useful discussion.

The paper had the financial support of the Russian Fund for
Fundamental Research (RFFI) \# 09-02-00551-a.

\end{document}